\documentclass{aa}  

\usepackage{txfonts}
\usepackage{newtxtext,newtxmath}
\usepackage[T1]{fontenc}
\usepackage[export]{adjustbox}
\usepackage{graphicx}	% Including figure files
\usepackage{amsmath}	% Advanced maths commands
\usepackage{amssymb}	% Extra maths symbols
\usepackage{subcaption}

\usepackage{etoolbox}
\usepackage{xcolor, ulem}

\providetoggle{long}
\settoggle{long}{true}

\usepackage[colorlinks,linkcolor=blue,anchorcolor=blue,citecolor=blue]{hyperref}
\newcommand{\sfrd}{$\Sigma_{\rm SFR}$\xspace}

\newcommand{\usfrd}{$\rm M_{\odot}\,yr^{-1}\,kpc^{-2}$}

\newcommand{\udif}{$\rm cm^2\,s^{-1}$}

\begin{document} 

    \title{Diffusion of cosmic-ray electrons in M~51 observed with LOFAR at 54~MHz}

   %\subtitle{Nearby galaxies in LoTSS-DR2}

   \author{V. Heesen\inst{1}
          \and
          F.~de~Gasperin\inst{1}
          \and
          S.~Schulz\inst{1}
          \and
          A.~Basu\inst{2}
          \and
          R.~Beck\inst{3}
          \and
          M.~Br\"uggen\inst{1}
          \and
          R.-J.~Dettmar\inst{4}
          \and
          M.~Stein\inst{4}
          \and
          L.~Gajovi\'c\inst{1}
          \and
          F.~S.~Tabatabaei\inst{5}
          \and
          P.~Reichherzer\inst{6}
          }

   \institute{Hamburger Sternwarte, University of Hamburg, Gojenbergsweg 112, 21029 Hamburg, Germany\\
              \email{volker.heesen@hs.uni-hamburg.de}
         \and
         Th\"uringer Landessternwarte, Sternwarte 5, 07778 Tautenburg, Germany 
         \and 
         Max-Planck Institut f\"ur Radioastronomie, Auf dem H\"ugel 69, 53121 Bonn, Germany
         \and
         Ruhr University Bochum, Faculty of Physics and Astronomy, Astronomical Institute, 44780 Bochum, Germany
         \and 
         School of Astronomy, Institute for Research in Fundamental Sciences, 19395-5531 Tehran, Iran
         \and
         Theoretical Physics IV: Plasma-Astroparticle Physics, Faculty for Physics \& Astronomy, Ruhr-Universit\"at Bochum, 44780 Bochum, Germany
       }

   \date{Accepted XXX. Received YYY; in original form ZZZ}
   
% \abstract{}{}{}{}{} 
% 5 {} token are mandatory
 
  \abstract
   {The details of cosmic-ray transport have a strong impact on galaxy evolution. The peak of the cosmic-ray energy distribution is observable in the radio continuum using the electrons as proxy.}
   {We measure the length that the cosmic-ray electrons (CRE) are transported during their lifetime in the nearby galaxy M~51 across one order of magnitude in cosmic-ray energy (approximately 1--10 GeV). To this end we  use new ultra-low frequency observations from the LOw Frequency ARay (LOFAR) at 54~MHz and ancillary data between 144 and 8350~MHz.}
   {As the the CRE originate from supernova remnants, the radio maps are smoothed in comparison to the distribution of the star formation. By convolving the map of the star-formation rate (SFR) surface density with a Gaussian kernel, we can linearise the radio--SFR relation. The best-fitting convolution kernel is then our estimate of the CRE transport length.}
   {We find that the CRE transport length increases at low frequencies, as expected since the CRE have longer lifetimes. The CRE transport length is $l_{\rm CRE}=\sqrt{4 D t_{\rm syn}}$, where $D$ is the isotropic diffusion coefficient and $t_{\rm syn}$ is the CRE lifetime as given by synchrotron and inverse Compton losses. We find that the data can be well fitted by diffusion,  where $D=(2.14\pm 0.13) \times 10^{28}$~\udif. With $D\propto E^{0.001\pm 0.185}$, the diffusion coefficient is independent of the CRE energy $E$ in the range considered.}
   {Our results suggest that the transport of GeV-cosmic ray electrons in the star-forming discs of galaxies is governed by energy-independent diffusion. }


   \maketitle
%
%-------------------------------------------------------------------

\section{Introduction}

The formation and evolution of galaxies is regulated by the accretion and ejection of matter.  The latter occurs in galactic winds, which are the result of star formation and subsequent supernovae. This so-called stellar feedback causes heating of gas and this hot ionised gas can thermally drive  winds \citep[see][for a recent review]{veilleux_20a}. Recently, it has become evident both from theoretical considerations \citep{breitschwerdt_91a,recchia_16a,yu_20a} and simulations \citep{salem_14a,pakmor_16a,jacob_18a} that cosmic rays can also drive winds on galactic scales. Indeed, cosmic ray-driven winds are thought to be more mass-loaded, containing the warm ionised medium with a much higher density \citep{girichidis_18a}. Also cosmic rays can potentially drive winds in galaxies, where the thermal gas alone is not sufficient to launch a wind \citep{everett_08a}. Hence, the influence of cosmic rays could lead to much more ubiquitous galactic winds that are also denser. This would have far-reaching consequences for galaxy evolution and the relation of galaxies with the circumgalactic medium \citep{tumlinson_17a}.

The details of cosmic-ray transport are paramount for properly incorporating them into models of galaxy evolution, with the various contributions from diffusion, advection and streaming to be included \citep{hopkins_20a}. Most of the understanding of cosmic-ray transport has come so far from our Galaxy, with observations such as those made with the AMS-02 experiment being able to precisely measure the composition and energy spectrum of cosmic rays \citep{aguilar_16a}. While there is no reason to believe that external galaxies should have any different behaviour, their observation provides us with a global view of a galaxy which potentially allows us to study the transport from the disc into the halo \citep{heesen_21a}. This is important as the  size of the halo into account, at the boundary of which cosmic rays are assumed to escape in the analysis of Galactic cosmic-ray transport, is a free parameter \citep{evoli_20a} resulting in a degeneracy between halo size and diffusion coefficient \citep{chan_19a}. Also, the gaseous thin disc and gaseous halo, which have very different magnetic field structures, are essential to study to get a meaningful description of cosmic ray transport. It has been hence suggested that a vertical gradient of cosmic rays in the halo can lead to streaming of cosmic rays with a velocity of the order of the Alfv{\'e}n speed along the magnetic field lines \citep{wiener_17a} which then can lead to a galactic wind \citep{uhlig_12a}. In contrast, the transport of cosmic rays in the disc may be more mediated by magnetic field irregularities that cascade down via turbulence from the injection scale to the gyro (Larmor) radius of $10^{11}~\rm cm$ for a 1~GeV CRE in a 10~$\upmu$G magnetic field \citep[the `extrinsic turbulence' picture, see][]{crocker_20a}. Then again, cosmic rays can generate their own turbulence and this `self confinement' picture is usually invoked in order to explain why cosmic rays are confined for a considerable time to the galaxy \citep{zweibel_13a}. The transport of cosmic-ray electrons may be particular difficult to model as the usual approximation of a continuous steady state distribution of cosmic ray sources can not be applied. The energy loss-time of very highly energetic electrons of about 1 TeV energy is only $\approx$$10^5$~yr in the interstellar medium. These particles may diffuse at given point in a galaxy only from a few nearby sources  such as supernova remnants.

%Hence, a study of cosmic ray transport in galaxies should focus on several aspects. The first aspect is to establish, whether cosmic-ray streaming or advection dominates in certain circumstances. If so, the advection or streaming speed can then be measured. The second aspect is to find situations where a random walk process dominates. In this case, the diffusion coefficient can be measured. The third aspect is the energy-dependence of the diffusion coefficient, which tells us something about the source of turbulence that cosmic rays encounter.

One way to study cosmic-ray transport of electrons is use to use radio continuum observations of nearby galaxies \citep[see][for an overview]{heesen_21a}. Radio continuum maps provide us with spatially resolved information via the synchrotron emission from cosmic-ray electrons (CRE) spiralling around magnetic field lines. These electrons as observed in the tens of MHz to tens of GHz, have energies of 1--10~GeV, corresponding to the peak of the cosmic-ray energy density. Hence, learning something about the CRE can help us us to study cosmic-ray transport for the dynamically most important part of the energy spectrum. The radio continuum images of galaxies show that the synchrotron emission is widely pervasive and only 10 per cent of the emission stems from supernovae remnants \citep{lisenfeld_00a}, the most likely source for cosmic-ray acceleration. As the diffuse radio continuum shows the presence of CRE that originate in supernova remnants, we can study cosmic-ray transport by comparing the distribution of massive star formation with that of the radio emission. In particular, one usually compares the distribution of the star-formation rate surface density, \sfrd, with the radio continuum intensity \citep[e.g.][]{murphy_08a,berkhuijsen_13a,tabatabaei_13b,heesen_14a,heesen_19a,vollmer_20a}. The radio continuum map is then the smoothed version of the \sfrd-map and the smoothing scale is then equated to the CRE transport length.

%In edge-on galaxies, radio haloes of a scale of 10~kpc become obvious and obviously cosmic rays have to be transported there from the star-forming disc \citep{wiegert_15a,heesen_18b}.

\citet{murphy_08a} compared 1365-MHz emission with 70-$\upmu$m far-infrared emission, a tracer for star formation, in a sample of 18 nearby galaxies. They found that the smoothing length-scale is about 1~kpc with some variation as function of the \sfrd-value with smaller lengths for higher \sfrd-values. At higher \sfrd-values, magnetic field strengths and so synchrotron losses are higher leading to reduced CRE number densities. Hence, this finding is consistent with CRE transport. \citet{vollmer_20a} studied galaxies at two frequencies, both at 1460 and 4850 MHz, in order to compare the smoothing length scale. They found that most galaxies appear to be diffusion dominated with a few having dominating cosmic ray streaming. \citet{heesen_19a} studied 3 galaxies, where it was found that they were consistent with diffusion. \citet{mulcahy_14a} used scale lengths at the edge of M~51 in order to measure the diffusion coefficient. A different approach in face-on galaxies is to use the radial radio spectral index distribution and model this with cosmic ray diffusion \citep{mulcahy_16a,doerner_22a}. This approach is in particular sensitive to the escape of cosmic rays, as the radio continuum spectrum becomes otherwise to steep. This escape can be more directly studied in edge-on galaxies that also allow us to measure the diffusion coefficient in radio haloes that are diffusion-dominated \citep{heesen_16a,heesen_19b,schmidt_19a,stein_19a}.

%What is so far lacking, is a clear demonstration of the cosmic-ray transport process undergoing in the galactic disc. Studies thus far are using two frequencies at most and while this can give an indication of whether diffusion or streaming dominates, it leaves rather room for interpretation. Also, the frequency range should be expanded in order to get the widest range of CRE energies and lifetimes. 

In this work, we expand these kind of studies to the lowest radio frequencies with new observations of the nearby galaxy M~51 (distance of $8.0$~Mpc) using the Low Frequency ARray \citep[LOFAR;][]{vanHaarlem_13a}. Observations of M~51 were already presented by \citet{heesen_19a} at 144~MHz using the high-band antenna (HBA) system. We significantly expand the frequency range to lower frequencies using the low-band antenna (LBA) system to obtain a map at 54~MHz. This map is the deepest and best resolved map of this source at this low frequency to date \citep{de_gasperin_21a}. By using ancillary data of frequencies up to 8350~MHz, we are now able to span a frequency range in excess of two orders of magnitude, corresponding to the more than an order of magnitude in cosmic-ray energy. We exploit these data in order to determine the so far most accurately measured relation between CRE transport length and lifetime. As we will show, our data are consistent with diffusion where the diffusion coefficient is energy independent and in good agreement with the canonical Galactic value.

This paper is organized as follows. In Section~\ref{s:data}, we describe our radio continuum data including the observations with LOFAR LBA. Section~\ref{s:measuring_the_cosmic_ray_transport_length} describes the methods we have employed to measure the CRE transport length. We we present our results in Section~\ref{s:results} and discuss their implications in Section~\ref{s:discussion}. We conclude in Section~\ref{s:conclusions}.

  %Our galaxy sample consists of 10 galaxies, some of their physical properties can be found in table \ref{tab:sample}.
\begin{figure*}
	\includegraphics[width=\textwidth]{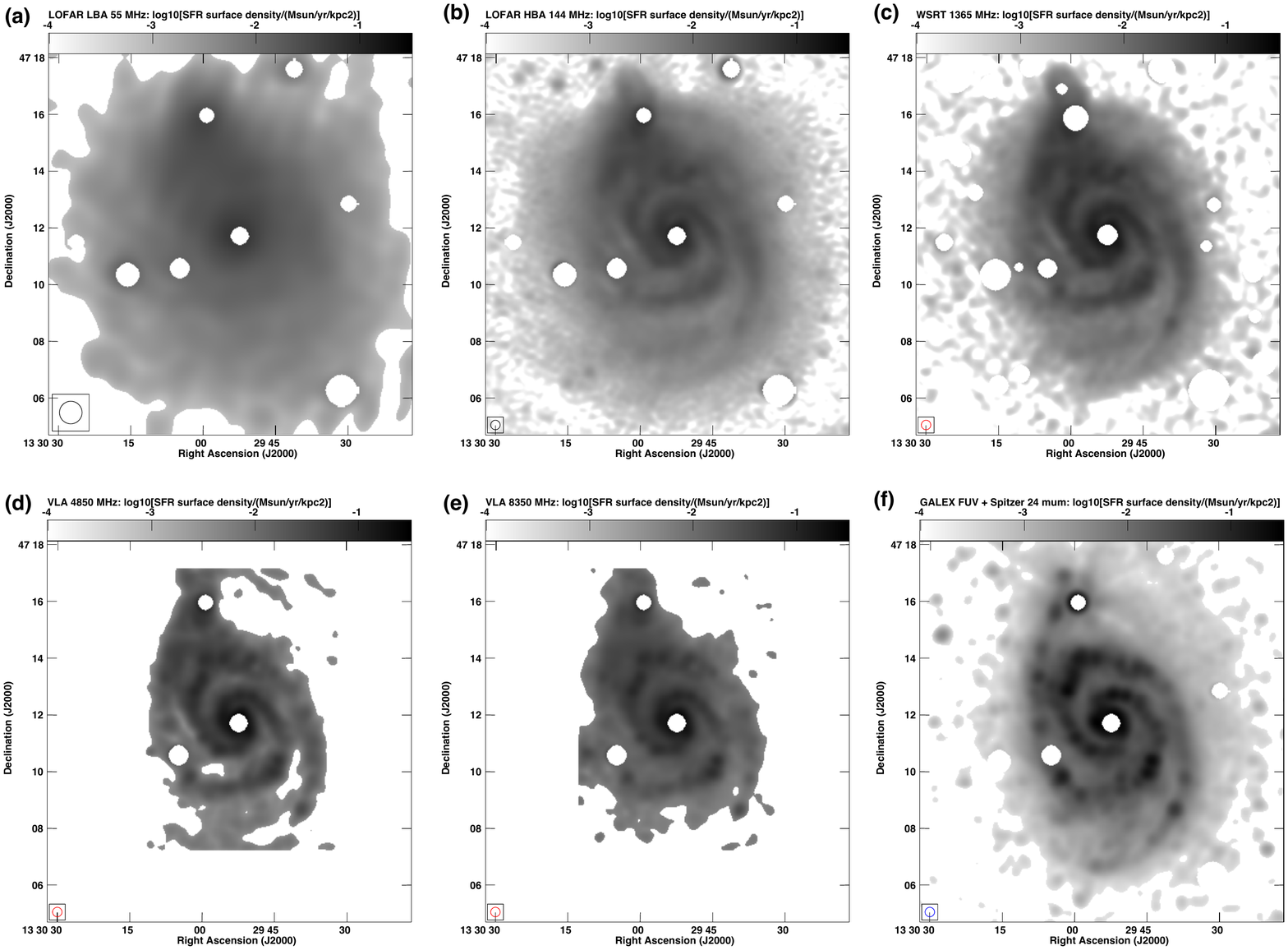}
    \caption{Radio continuum maps converted to star-formation rate surface density (\sfrd). The scaling is logarithmic between $10^{-4}$~\usfrd and $3\times 10^{-1}$~\usfrd. In the top row are LOFAR LBA 54~MHz (panel a), LOFAR HBA 144~MHz (b), and WSRT 1365~MHz (c). In the bottom row are VLA+Effelsberg 4850 MHz (panel d) and 8350~MHz (e). In the bottom right (panel f), the hybrid \sfrd-map is shown for comparison. The angular resolution is 20~arcsec FWHM for all maps, except for LOFAR LBA where it is 47~arcsec. The radio continuum maps appear to be the smoothed versions of the hybrid \sfrd-map as expected for CRE diffusion.}
    \label{fig:maps}
\end{figure*}

\section{Data}
\label{s:data}

\subsection{LOFAR LBA}
\label{ss:lofar_lba}

We observed with LOFAR LBA between 2017 and 2019 as part of the LOFAR  LBA Sky Survey \citep[LoLSS;][]{de_gasperin_21a} in the frequency range 42--66 MHz. The LBA stations were used in LBA\_OUTER mode, meaning that the primary beam has a size of 4$\degr$ full width at half-maximum (FWHM), closely matched to the HBA primary beam. The $(u,v)$ data include baselines of up to 100~km in length, so that a nominal resolution of 15~arcsec can be reached. The final data release of LoLSS will reach this resolution at an rms noise level of $1~\rm mJy\,beam^{-1}$. This will entail a direction-dependent calibration technique which is currently developed and explored \citep{de_gasperin_20a}. This paper uses the preliminary LoLSS data release, which has a resolution of 47~arcsec and a median rms noise of $5~\rm mJy\,beam^{-1}$ covering 740~deg$^2$ around the HETDEX field \citep{de_gasperin_21a}. The actual noise of our field is indeed slightly better with $3.3~\rm mJy\,beam^{-1}$. This means our sensitivity for extended emission on the scale of one synthesized beam is comparable to the final data release.

The data were first calibrated in standard fashion with the {\sc PiLL} pipeline \citep{de_gasperin_19a}. As stated before, only direction-independent calibration effects were taken into account. The imaging was performed with {\sc WSClean} \citep{offringa_14a} with Briggs weighting $-0.3$ and multi-scale cleaning. An outer ($u,v$)-range of 4500$\lambda$ was used in order to limit direction-dependent ionospheric errors. The 95 direction-dependent calibrated images were combined into a single large mosaic. This mosaic formed the basis for our analysis presented here.

\subsection{Other data}
The LOFAR HBA data at 144~MHz was obtained from the LoTSS survey data release 2 \citep{shimwell_22a}, where we took the 20-arcsec map presented in \citet{heesen_22a}. The 1365-MHz map is from observations with the Westerbork Synthesis Radio Telescope (WSRT) presented by \citet{braun_07a}. We also used a 4850-MHz map and 8350-MHz map from \citet{fletcher_11a}, who combined observations with the Very Large Array (VLA) with single-dish data using observations with the 100-m Effelsberg telescope. The largest angular scale that the WSRT can observe at 1365~MHz is 25~arcmin when one considers base lines of larger than 27~m \citep{heesen_14a}. As M~51 has an angular extent of 14~arcmin in the radio continuum, we do not have to correct for missing zero-spacings of our interferometric radio data. For the VLA observations a correction was made merging the VLA and Effelsberg maps in the image space \citep{fletcher_11a}.  The hybrid SFR maps are taken from \citet{leroy_08a}, who use a combination of {\it GALEX} far--UV 156-nm  and {\it Spitzer} mid-IR 24-$\upmu$m data. We applied a correction for thermal emission using H\,$\alpha$ data which were corrected for absorption by dust using a combination of {\it Spitzer} 70- and 160-$\upmu$m maps \citep{tabatabaei_07b} which we will be presented by Tabtabaei et al. (in prep). The integrated thermal flux density is 55~mJy at 1400~MHz, which is in fair agreement with the 5\,\% thermal fraction reported by \citet{tabatabaei_17a} from radio spectral energy distribution fitting.

Maps were transformed to the same coordinate system and convolved with a Gaussian to a resolution to 20~arcsec; the 54-MHz map was left at native resolution. Strong contaminating sources including the nuclei of both M~51a and M~51b were masked, as as were a few compact sources within the disc of M~51. These have sizes of $\lesssim$ 10~milli arcseconds \citep{rampadarath_15a}, hence are either supernova remnants within M~51 or active galactic nuclei in the background. In order to ease the comparison with the hybrid star-formation rate surface density map, we converted the maps into  radio-\sfrd, $(\Sigma_{\rm SFR})_{\rm RC}$. We used Condon's \citep{condon_92a} linear relation between radio continuum luminosity and star-formation rate. We then converted the radio continuum intensity into $(\Sigma_{\rm SFR})_{\rm RC}$ as described in \citet{heesen_14a} and \citet{heesen_19a} assuming a radio spectral index of $-0.8$. This radio spectral index is in good agreement with the integrated radio continuum spectrum of M\,51 \citep{mulcahy_14a}.

These steps were carried out in the Astronomical Image Processing System ({\sc aips}), where we made use of the {\sc parseltongue} {\sc python} interface \citep{kettenis_06a}.

\begin{figure*}
	\includegraphics[width=\textwidth]{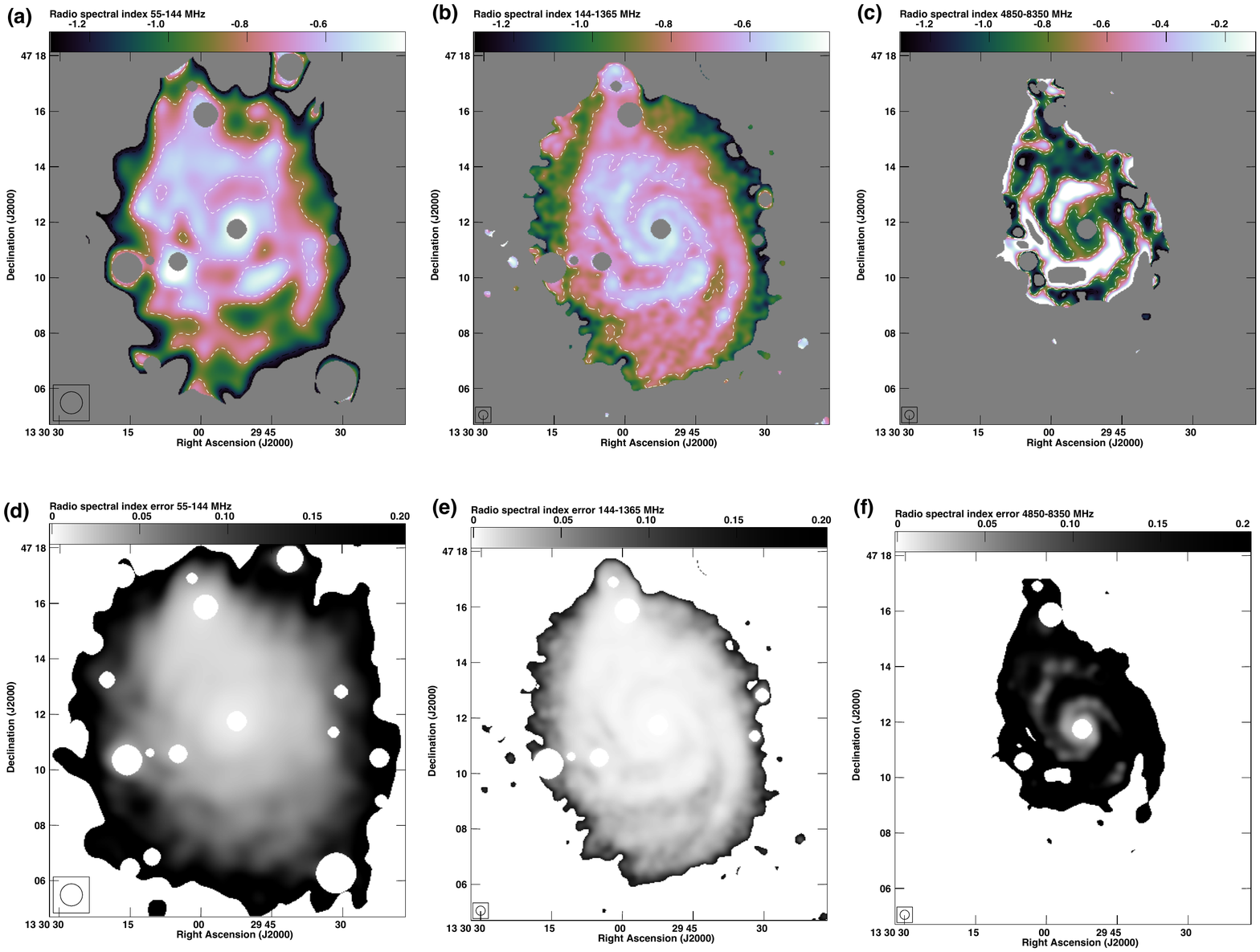}
    \caption{Radio spectral index distribution. The top row shows from left to right the non-thermal radio spectral index between 54 and 144~MHz (panel a), between 144 and 1365~MHz (panel b), and between 4850 and 8350~MHz (panel c). Contours are at $-1.2$, $-0.85$ and $-0.65$. The bottom row (panels d--f) shows the corresponding radio spectral index error. The angular resolution is 47~arcsec for the maps between 54 and 144 MHz and 20~arcsec otherwise. The size of the synthesised beam is shown in the bottom-left corner of each panel.}
    \label{fig:spix}
\end{figure*}

\section{Measuring the cosmic-ray transport length}
\label{s:measuring_the_cosmic_ray_transport_length}
\subsection{Morphology of the radio continuum maps}
\label{subsec:morphology}

In Fig.~\ref{fig:maps} we present the radio continuum maps at various frequencies. We can already visually see a clear trend of a decreasing smoothness of the radio continuum emission with increasing frequency (Fig.~\ref{fig:maps}a--e) when compared with the hybrid \sfrd-map (Fig.~\ref{fig:maps}f). In particular, in the outskirts the galaxy appears to shrink and the arm--interarm contrast increases. This is borne out in the radio spectral index maps shown in Fig~\ref{fig:spix}, where in the outskirts of the galaxy the radio spectral index is particularly steep with $\alpha<-0.9$ (areas in green and blue) both in the 54--144 MHz (Fig.~\ref{fig:spix}a) and 144--1365 (Fig.~\ref{fig:spix}b) radio spectral index maps. At higher frequencies, in the 4850--8350~MHz radio spectral index map (Fig.~\ref{fig:spix}c), the outskirts become invisible due to spectral ageing, but the inter-arm regions have a steep radio spectrum. In contrast, at all frequencies the radio spectrum is consistent with young CRE in the spiral arms, where the radio spectral index is $\alpha>-0.6$.

The radio spectral index shows areas where the radio continuum spectrum is flatter than expected for an injection spectral index of $\alpha\approx -0.5$, in particular in the spiral arms. At low frequencies $\lesssim$144~MHz, this may be caused by thermal absorption where H\,{\sc ii} regions become optically thick. Alternatively, the low-energy CRE spectrum may flatten due to the combination of ionisation and bremsstrahlung losses \citep{basu_15a}. With the new 54-MHz data this can be explored in more detail in the future. At higher frequencies, we also find such flat radio continuum spectra in the spiral arms, which are prominent in the 4850--8350~MHz radio spectral index map. Since we corrected for thermal emission and the emission would have to be almost 100\,\% thermal, we attribute this to uncertainties in observations of the radio continuum emission. For instance, a small decrease of radio emission at 4850~MHz which can be caused by the merging of interferometric and single-dish data \citep[VLA and Effelsberg data in the case of][]{fletcher_11a}. Such a deviation is suggested by the radio continuum spectrum of the integrated flux density \citep[see fig.~2 in][]{kierdorf_20a}. 

\subsection{Radio--star formation rate relation}
\label{sec:plotting}

We first analyse the spatially resolved radio continuum--star-formation rate (radio--SFR) relation.  We  convolve the maps to a spatial resolution of 1.2~kpc corresponding to $30.1$~arcsec (FWHM) and bin them to a pixel size of $1.2\times 1.2~\rm kpc^2$ ($1.8\times 1.8~\rm kpc^2$ for 54~MHz). Our analysis was done using the software \texttt{radio--pixel--plots} ({\sc rpp})\footnote{\url{github.com/sebastian-schulz/radio-pixel-plots}}. We take two radio continuum maps at different frequencies and the \sfrd-map, calculate radio and hybdrid \sfrd-values and radio spectral indices. Then we fit the radio continuum \sfrd-values as function of the hybrid \sfrd-values with following function:
 \begin{equation}
    \log_{10} \left[ (\Sigma_{\text{SFR}})_{\text{RC}} \right] = a \log_{10} \left[ (\Sigma_{\text{SFR}})_{\text{hyb}} \right] + b. 
    \label{eq:loglog_fit} 
 \end{equation}
For the fitting we use the orthogonal distance regression ({\sc odr}) package from the {\sc SciPy} {\sc python} library, which uses a modified Levenberg--Marquardt algorithm. Its strength is dealing with both $x$ and $y$ errors even in case of large errors. Even though the measurement uncertainties in our radio and hybrid SFR data are small, the conversion to \sfrd-values has an error of  approximately 50\,\% \citep{leroy_12a}.

For the uncertainties we consider both the map rms noise  and a calibration error of 5\,\% \citep{heesen_14a}. After the resolution is converted to 1.2 kpc we apply a $3 \sigma$ cutoff. At the same time a second cutoff based on the radio spectral index is applied. Any pixel with an associated spectral index $\alpha > -0.5$ is flagged, will be plotted as an outlier and also excluded from the fitting data. The motivation for this is to exclude pixels where the radio continuum emission is dominated by thermal emission and thermal absorption \citep{basu_15a}. In practice, the number of points clipped is very small for frequencies below $1.4$~GHz, because we expect the radio continuum to be dominated by non--thermal emission. \citep{tabatabaei_17a}

%Let $X[i]$ denote the value of the $i$--th pixel in one coarse map and let $\sigma_{\text{noise}}$ be the noise in that map, then the total standard deviation of $X[i]$ is
%\begin{align}
%    \sigma_i = \sqrt{\sigma_{\text{noise}} ^2 + \left( X[i] \cdot 0.05  \right)^2} \,.
%\end{align}
%Note that since we need all three values (two radio and one SFR) any pixel that is below the 3 $\sigma$ threshold will be removed from all maps.\\

%Finally the coarse radio map is converted into SFR surface densities using the Condon--relation (Eq. \ref{eq:condon_final}) and then mapped onto the SFR map. In a double logarithmic plot they are expected to follow a linear relationship given by

We assume  that our errors are, in good approximation, described by a normal distribution and that the ratio of estimated errors for the radio and hybrid \sfrd is reasonably well known. Such Gaussian statistics are a good approximation for hybrid \sfrd as demonstrated by \citet{leroy_12a} with a statistical uncertainty of $0.13$~dex and a systematic uncertainty of up to $0.2$~dex; this statistics can be reasonably extended to radio \sfrd with a similar accuracy \citep{heesen_14a,heesen_19a}. In that case {\sc odr} works better than the ordinary least--squares regression as it has a smaller bias, lower variance of parameters and a smaller mean square errors of parameters \citep{boggs_89a}. Best-fitting parameters are presented in Table~\ref{tab:fitting}.

%The final radio--SFR plot contains the resulting best fit as well as the Condon--relation which always corresponds to the $y=x$ line because we used it to convert the radio data.
%\begin{figure}[h]
%    \centering
%    \includegraphics[scale=1.2]{n5194_l1200_pixel.pdf}
%    \caption{Radio--SFR plot of NGC 5194. The resulting value for the slope is $0.52\pm 0.02$. The points' different colors represent different ages of CREs based on their spectral indices: red are the youngest, blue the oldest.}
%    \label{fig:pixelplot}
%\end{figure}

\subsection{Smoothing experiment}
\label{sec:smoothing}
\citet{berkhuijsen_13a} and \citet{heesen_14a} showed that it is possible to linearize the spatially resolved radio--SFR relation by convolving the hybrid \sfrd-maps with a Gaussian kernel. The choice of kernel is motivated by our assumption that diffusion and not advection (or streaming) is the mechanism for CRE transport in the discs of galaxies. Note, however, that in the haloes of galaxies advection may be the dominating transport mode \citep[see e.g.][for a recent example]{heald_22a}. \citet{murphy_08a} and \citet{vollmer_20a} have shown that both exponential and Gaussian kernels work equally well in these smoothing experiments, so that we can not simply distinguish the transport mode this way. Hence, we can only later motivate the choice of the Gaussian kernel after we investigated the cosmic-ray transport length as function of CRE lifetime. We also assume the diffusion is cylindric (2D) isotropic with a preferred diffusion in the disc plane due to the large-scale ordered magnetic field. Hence, we use an elliptical kernel with a standard deviation $\sigma_x$ along the major axis and similar $\sigma_y$ along the minor axis, where we define:
\begin{equation}
    \sigma_x = \sigma_{xy},~~~\sigma_y=\cos(i)\sigma_{xy},
    \label{eq:elliptical_kernel}
\end{equation}
where $i=20\degr$ is the inclination angle. This is of course a simplification as CR diffusion is anisotropic due to the anisotropy of the eddy scales along the orientation of the `local background field', which is defined to have a scale that is larger than the eddy scale \citep[see e.g.\ Sect.~2.1 in][]{zhang_20a}. The relevant local field is less regular and stronger than the kpc-scale ordered field found in M\,51 and other galaxies. Polarisation observations with spatial resolutions between the outer scale of the turbulence ($\approx$50 pc) and the transport length ($\approx$2 kpc) reveal the ordered local fields that could lead to anisotropy. With the best resolution in \citet{fletcher_11a} of $\approx$150 pc the field pattern is still ordered with spiral shape, so that anisotropic diffusion is indeed probable.

We define the CRE transport length $l_{\text{CRE}}$ as half of the FWHM of the Gaussian kernel.  Following \citet{heesen_14a}, we set the transport length equal to the diffusion length and the conversion from FWHM is:
\begin{align}
    l_{\text{CRE}} = \frac{1}{2} ( 2 \sqrt{2 \ln 2}) \sigma_{xy} \approx 1.177  \, \sigma_{xy}\,,
\end{align}
where $\sigma_{xy}$ is the standard deviation of the Gaussian kernel.\\

The convolution is done using the \texttt{convolve\_fft} method from the {\sc AstroPy} library, where the kernel is defined using {\sc AstroPy}'s \texttt{Gaussian2DKernel} method and takes standard deviations $\sigma_{xy}$ as an input. In order to find the best-fitting kernel size that linearizes the radio--SFR we define the following function $C: \sigma_{xy} \mapsto a$ as:
\begin{enumerate}
    \item convolve the hybrid \sfrd-map with a Gaussian kernel of size $\sigma_{xy}$;
    \item convert radio maps to radio \sfrd-maps;
    \item bin maps to a pixel size of $1.2\times 1.2~\rm kpc^2$;
    \item clip maps using 3$\sigma$ and $\alpha$ thresholds;
    \item perform the double logarithmic linear fit to measure the slope $a$.
\end{enumerate}
%Steps 2, 4 and 5 work exactly as described in the section \ref{sec:plotting}. 
%In every iteration step 3 starts with the full coarse maps, because the absolute calibration errors for  the hybrid SFR map may change after each convolution.  The noise value $\sigma_{\text{noise}}$ for the hybrid SFR map is not updated although it could change as well. The changes are probably small and would lead to a minimal increase of $\sigma_{\text{noise}}$ which is why we leave it unchanged. In practice, the number of points after clipping does rarely change when compared to section \ref{sec:plotting}.\\

%\begin{figure}[h]
%    \centering
%        \includegraphics[scale=1.2]{NGC_5194_l1200_pixel_conv.pdf}
%    \caption{Smoothed radio--SFR plot of M51.}
%    \label{fig:convolution_example}
%\end{figure}

%, because diffusion is expected to modify the spatially resolved RC--SFR relation such that $(\Sigma_{\text{SFR}})_{\text{RC}} = ((\Sigma_{\text{SFR}})_{\text{hyb}} )^a$. 

With this definition linearizing the radio--SFR relation is equivalent to finding a solution for $C(\sigma_{xy})-1=0$. {\sc rpp} uses the iterative root finder \texttt{scipy.optimize.fsolve} from {\sc SciPy}, which is based on the  \textsc{fortran} library {\sc minpack}. We use two adjacent frequencies in order to calculate the radio spectral index. For every frequency, except for the lowest and the highest one, we get thus two measurements of the CRE transport length. The difference between these measurements allows us to estimate the error of $0.15$~kpc, which is larger than the statistical fitting error of $\lesssim$$0.07$~kpc. The best-fitting CRE transport lengths are presented in Table~\ref{tab:length}.

We note that there is some freedom of choosing the shape of the diffusion kernel. Numerical studies on CR transport fit diffusion coefficients using the second moment of their spatial distribution \citep{qin_09a,wang_19a,xu_13a,snodin_16a,seta_18a}. Based on magnetohydrodynamical turbulence simulations \citet{sampson_22a}
find that super-diffusion is ubiquitous in the ISM: super-diffusion would result in kernel shapes that have a narrower core and more extended wings than a Gaussian. The width of the distribution would also be modulated by the density distribution. Our data which represents just one snapshot in time does does not allow for a distinction between diffusive and super-diffusive regimes. However, this is something that could be explored in future work.

%except for the combination of 4850 and 8350~MHz, which we omit in favour of 1365--4850 MHz and 1365--8350~MHz, respectively. The reason is that the frequency coverage is wider so that the spectral index becomes more accurate.

%Provided one chooses a reasonable starting value for $\sigma_{xy}$, the algorithm will usually converge in less than 15 iterations.
%An example of the resulting radio--SFR plot is presented in figure \ref{fig:convolution_example}.

\begin{figure*}[!htp]
	\begin{subfigure}[t]{0.03\linewidth}
        \textbf{(a)}    
    \end{subfigure}
    \begin{subfigure}[t]{0.47\linewidth}
            \includegraphics[width=\linewidth,valign=t]{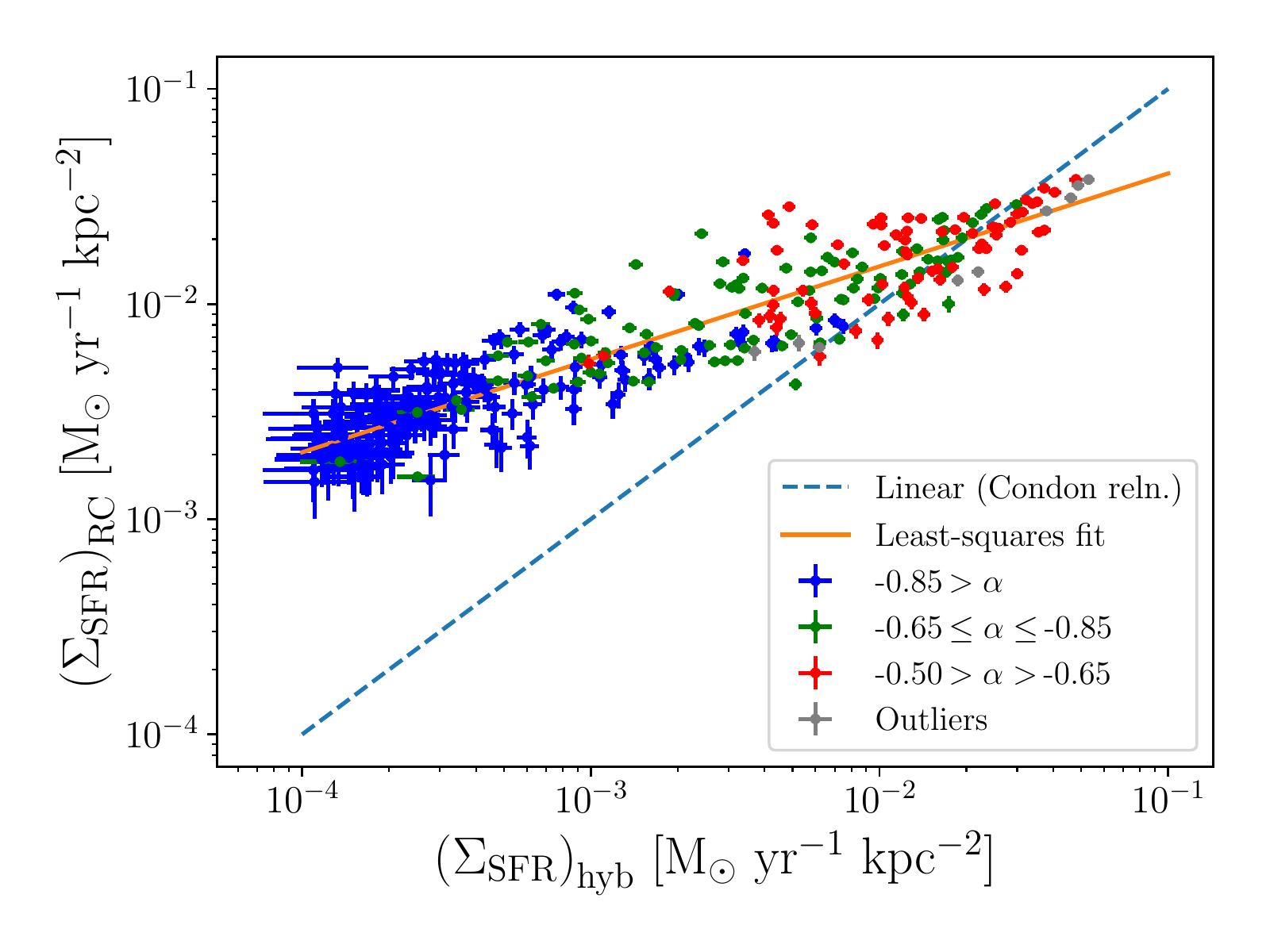}
    \end{subfigure}
    \begin{subfigure}[t]{0.03\linewidth}
    \textbf{(b)}    
    \end{subfigure}
    \begin{subfigure}[t]{0.47\linewidth}
            \includegraphics[width=\linewidth,valign=t]{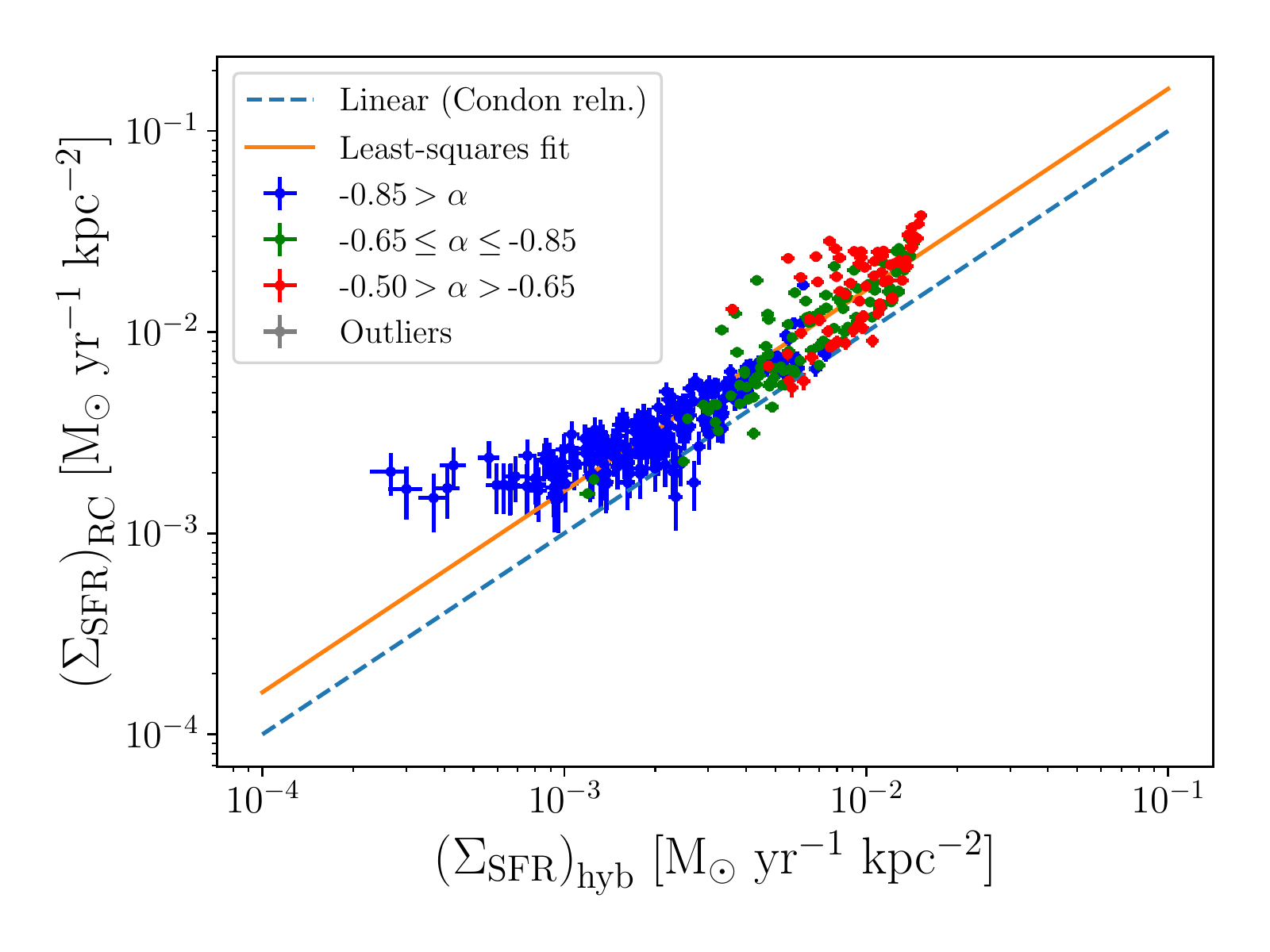}
    \end{subfigure}
    \\
    \begin{subfigure}[t]{0.03\linewidth}
        \textbf{(c)}    
    \end{subfigure}
    \begin{subfigure}[t]{0.47\linewidth}
            \includegraphics[width=\linewidth,valign=t]{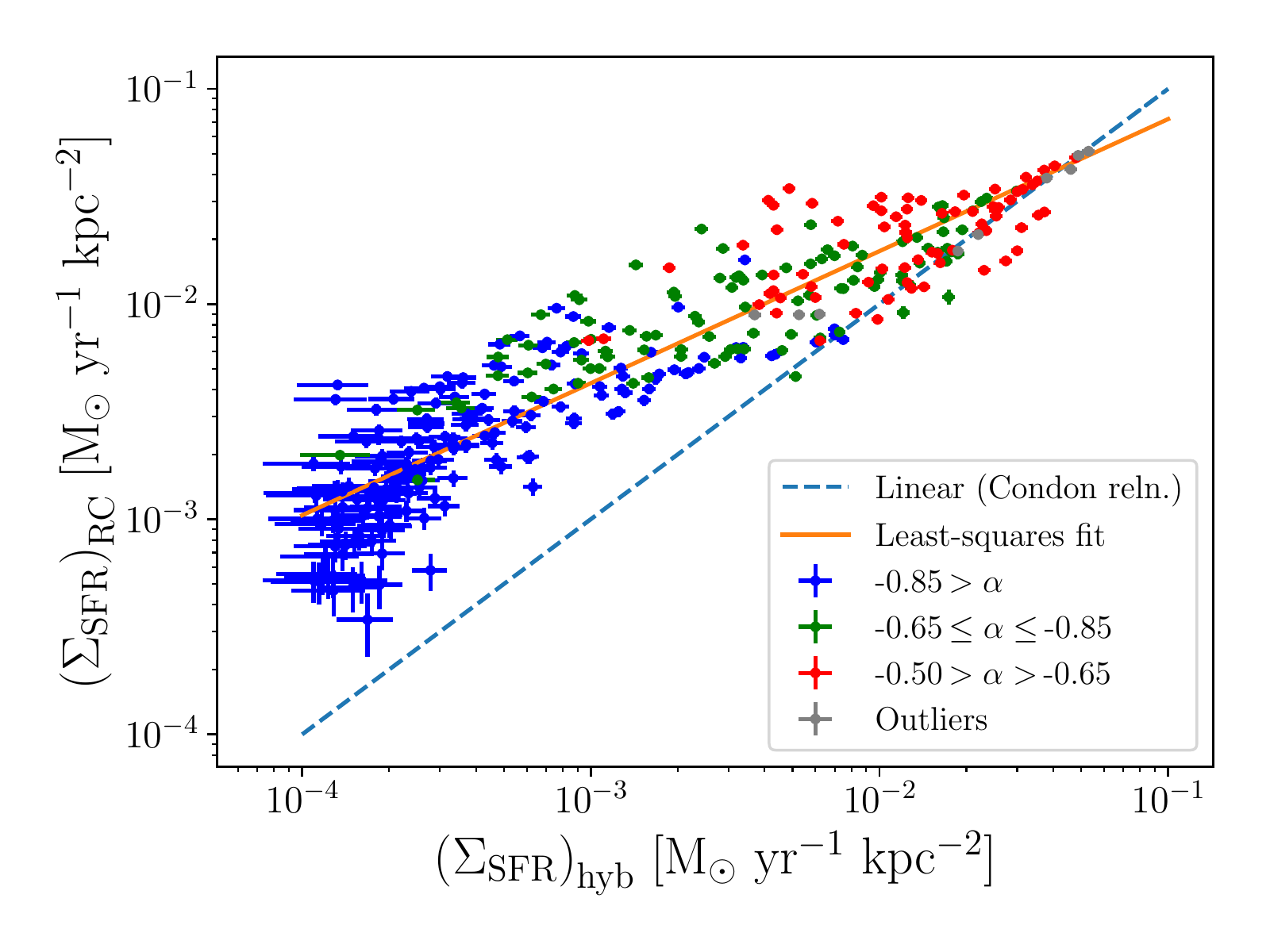}
    \end{subfigure}
    	\begin{subfigure}[t]{0.03\linewidth}
        \textbf{(d)}    
    \end{subfigure}
    \begin{subfigure}[t]{0.47\linewidth}
            \includegraphics[width=\linewidth,valign=t]{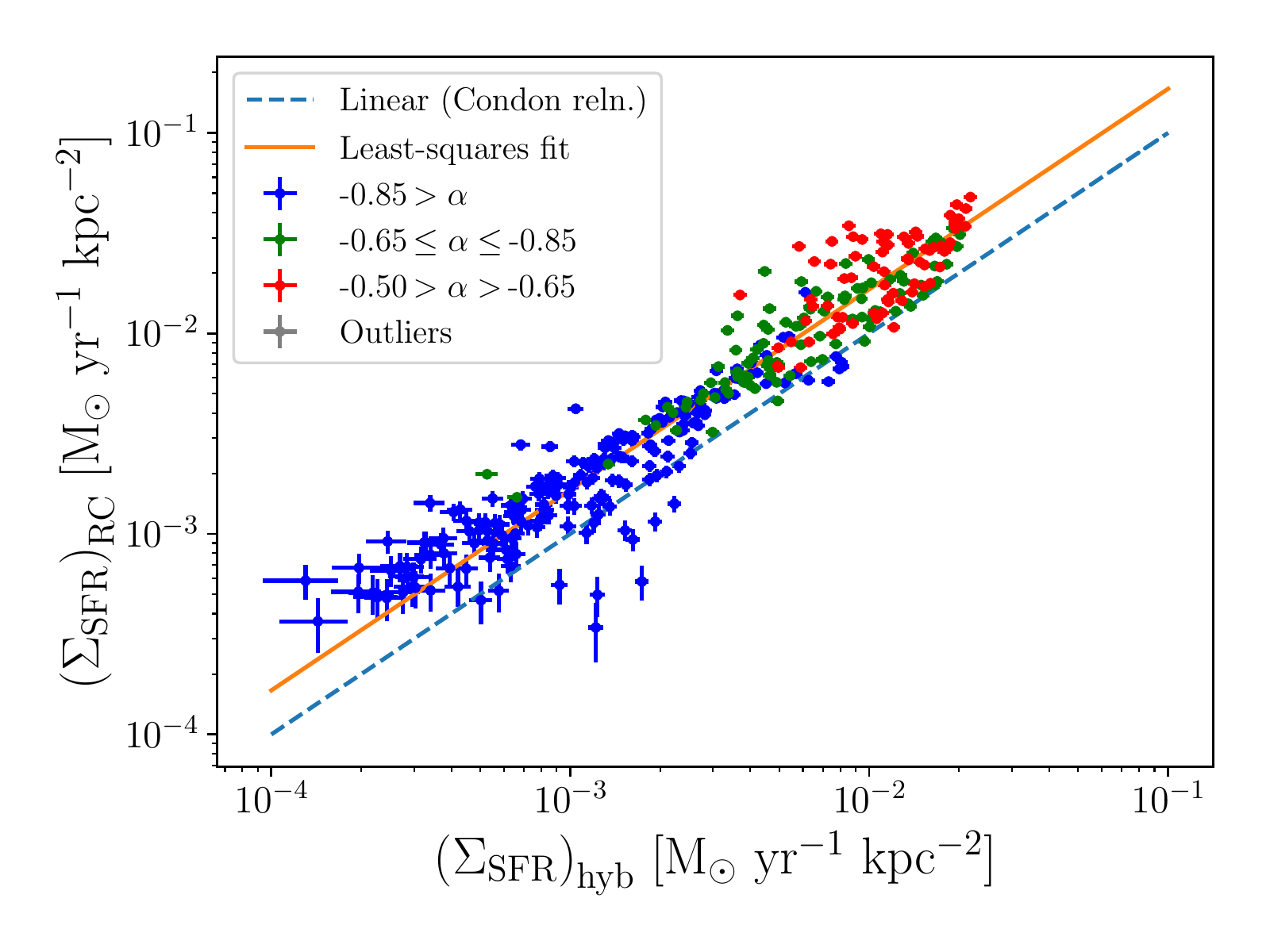}
    \end{subfigure}
    \\
    \begin{subfigure}[t]{0.03\linewidth}
        \textbf{(e)}    
    \end{subfigure}
    \begin{subfigure}[t]{0.47\linewidth}
            \includegraphics[width=\linewidth,valign=t]{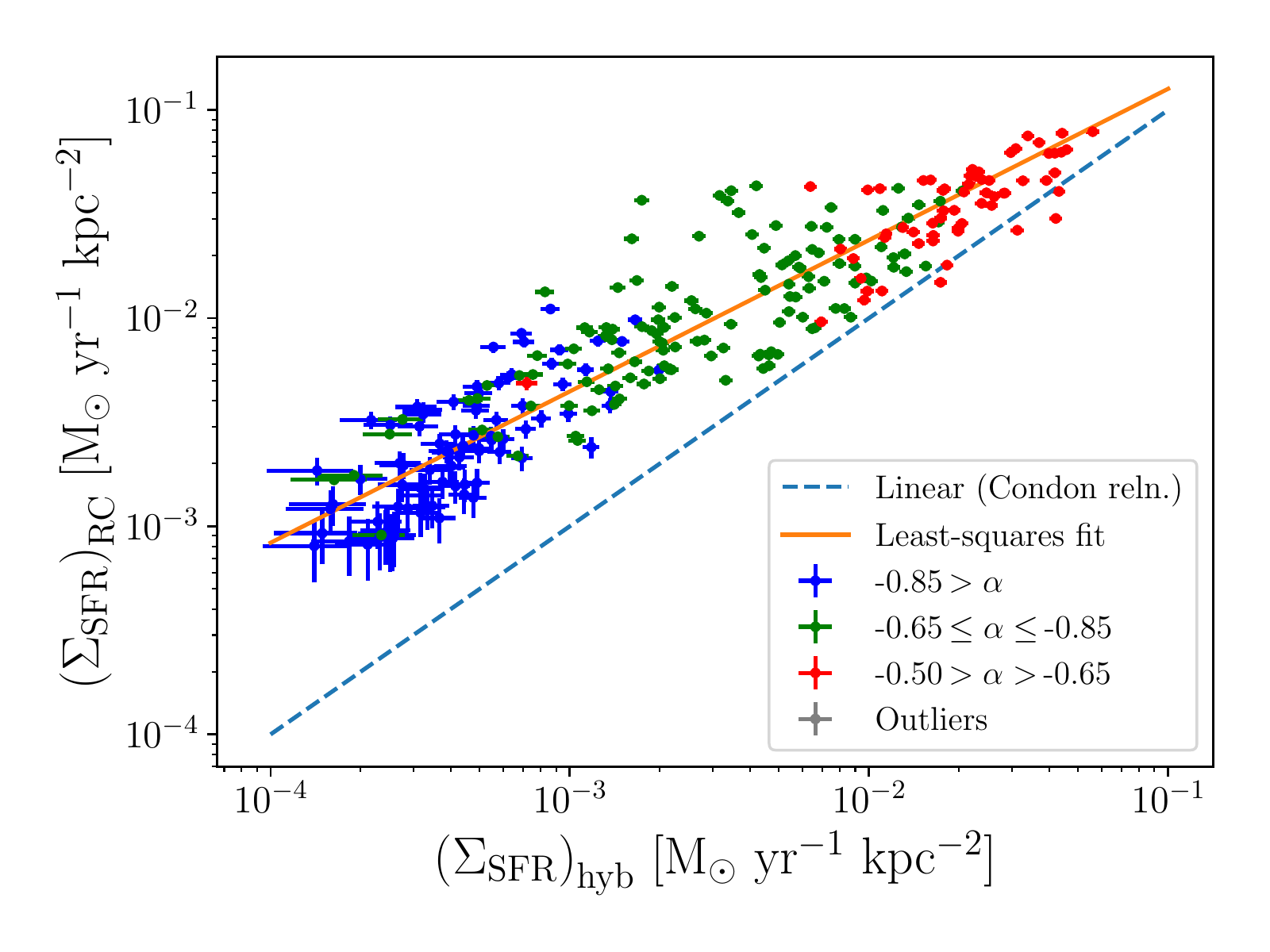}
    \end{subfigure}
    	\begin{subfigure}[t]{0.03\linewidth}
        \textbf{(f)}    
    \end{subfigure}
    \begin{subfigure}[t]{0.47\linewidth}
            \includegraphics[width=\linewidth,valign=t]{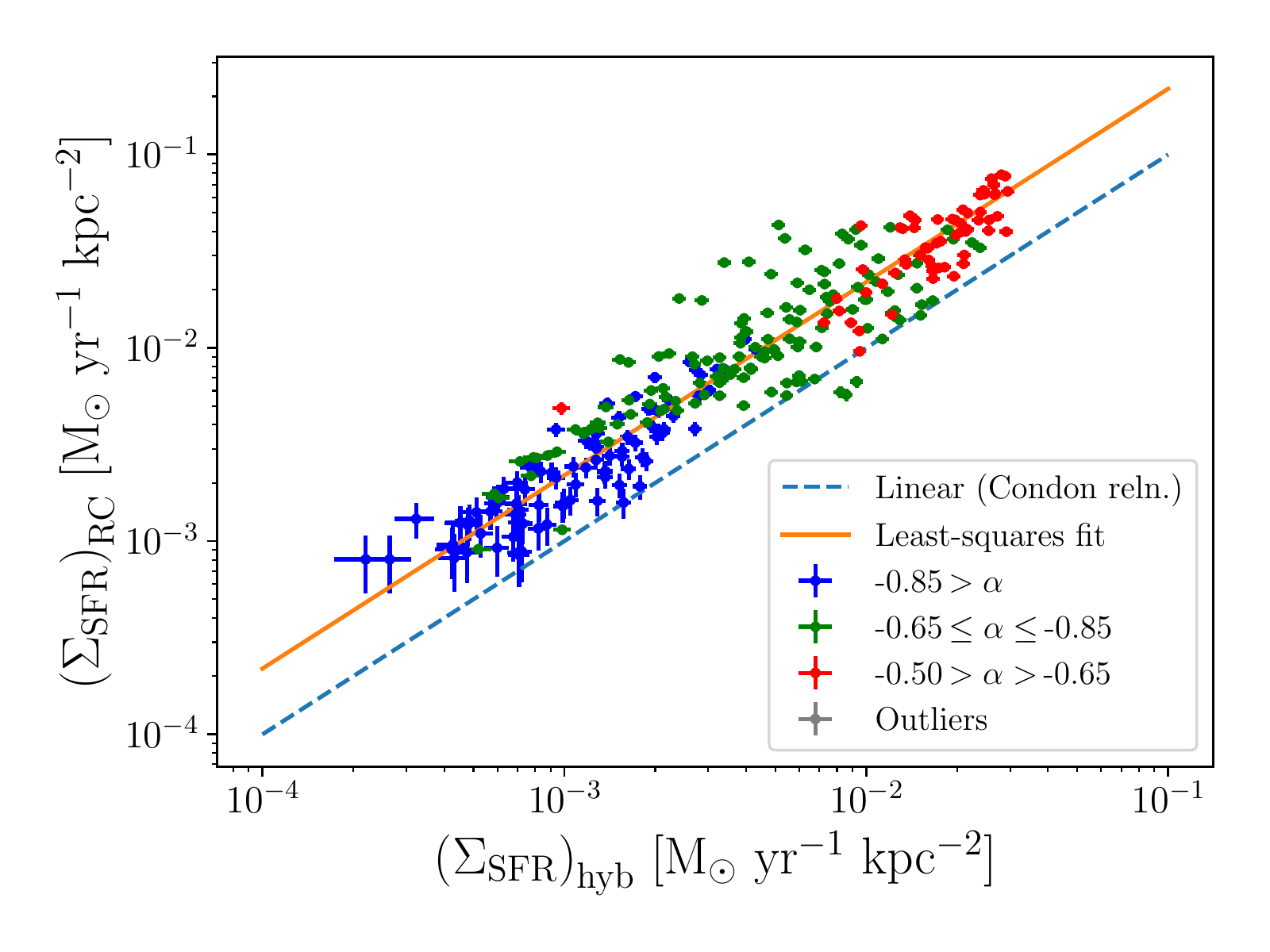}
    \end{subfigure}
    \caption{Cosmic-ray smoothing experiment. Panels show the spatially resolved radio--SFR relation prior to the smoothing (left panels) and following the smoothing of the \sfrd-maps (right panels). The vertical axis shows the radio continuum \sfrd-values converted with a constant factor from the radio continuum intensity. The horizontal axis shows the hybrid \sfrd-values. Each data point represents a pixel of $1.2\times 1.2~\rm kpc^2$. Data points are colour-coded with the non-thermal radio spectral index: data points with $\alpha<-0.85$ in blue, data points with $-0.85<\alpha<-0.65$ in green and data points with $-0.65<\alpha<-0.50$ in red. Data points with $\alpha>-0.5$ are shown as outliers and not considered for the data analysis; they are shown in grey. The best-fitting relation is shown as solid line and the Condon relation is shown as dashed line. Smoothing of the \sfrd-maps allows us to reach a slope of 1 for the radio--SFR relation. The vertical displacement is deviation from the integrated radio--SFR relation. From top to bottom, frequencies are 54~MHz (in combination with 144~MHz, panels a and b), 144~MHz (in combination with 54 MHz, panels c and d) and 1365~MHz (in combination with 144~MHz, panels e and f).}
    \label{fig:pixel}
\end{figure*}

\begin{figure*}
	\begin{subfigure}[t]{0.03\linewidth}
        \textbf{(a)}    
    \end{subfigure}
    \begin{subfigure}[t]{0.47\linewidth}
            \includegraphics[width=\linewidth,valign=t]{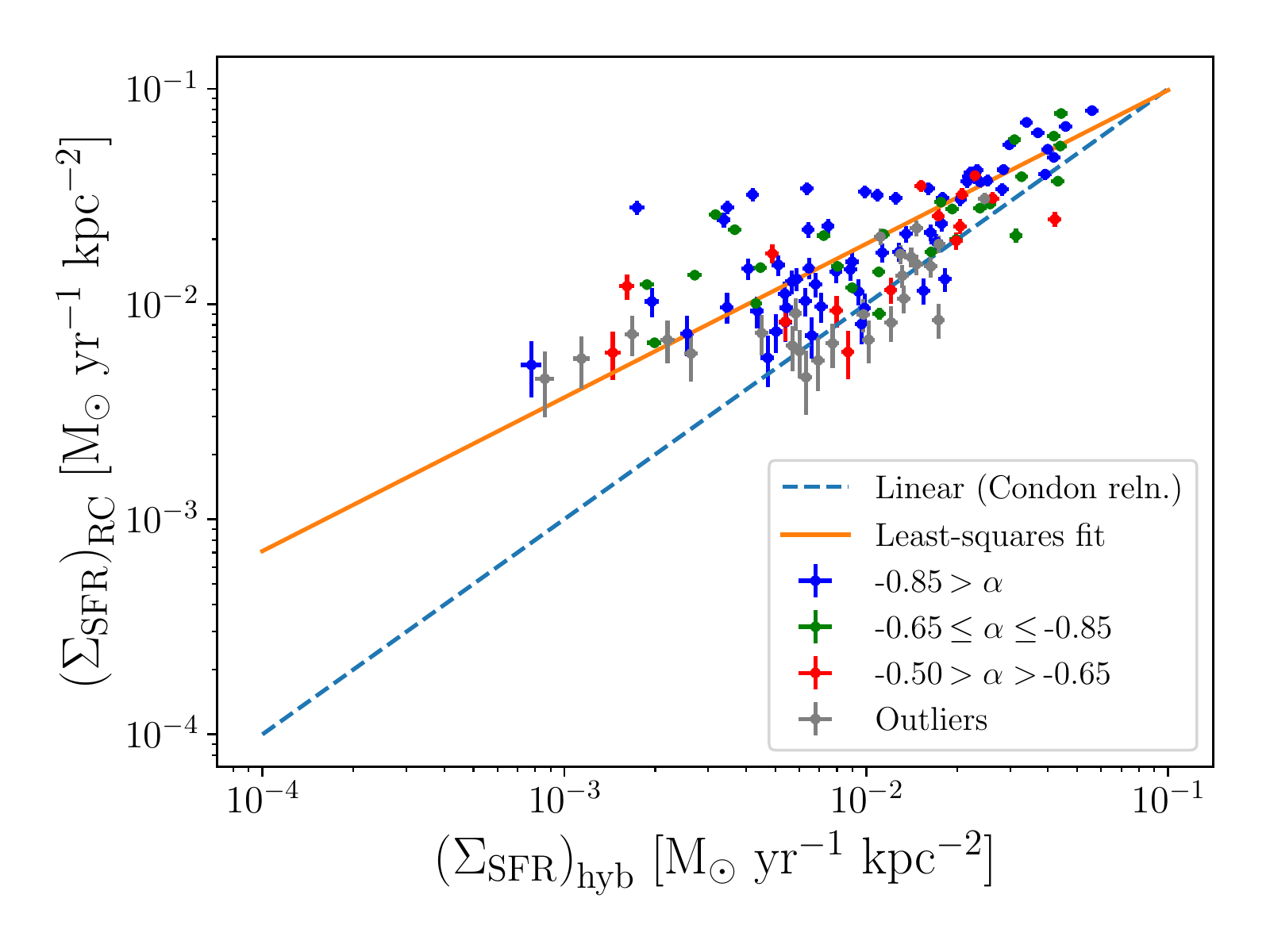}
    \end{subfigure}
    \begin{subfigure}[t]{0.03\linewidth}
        \textbf{(b)}    
    \end{subfigure}
    \begin{subfigure}[t]{0.47\linewidth}
            \includegraphics[width=\linewidth,valign=t]{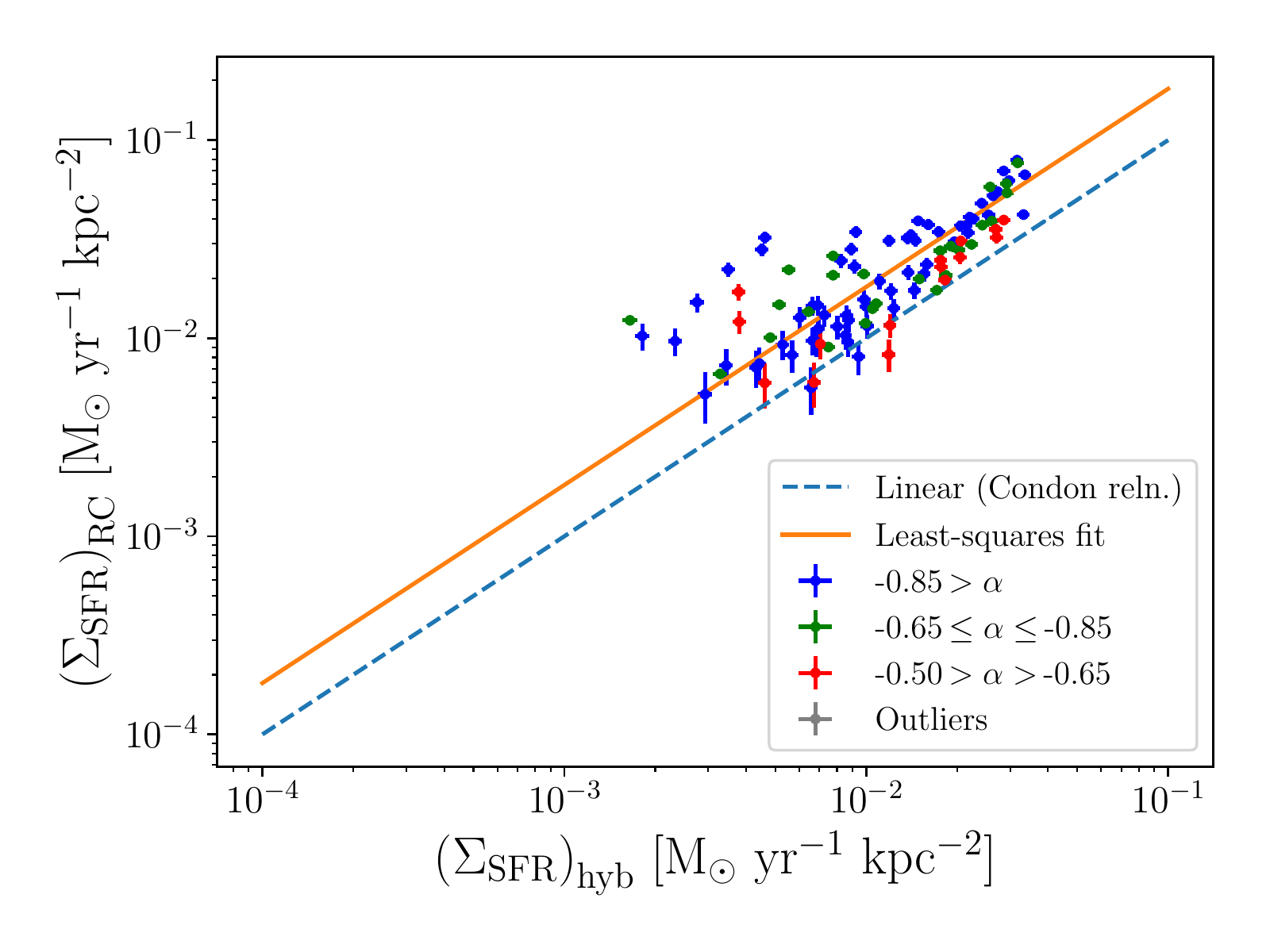}
    \end{subfigure}
    \\
    \begin{subfigure}[t]{0.03\linewidth}
        \textbf{(c)}    
    \end{subfigure}
    \begin{subfigure}[t]{0.47\linewidth}
            \includegraphics[width=\linewidth,valign=t]{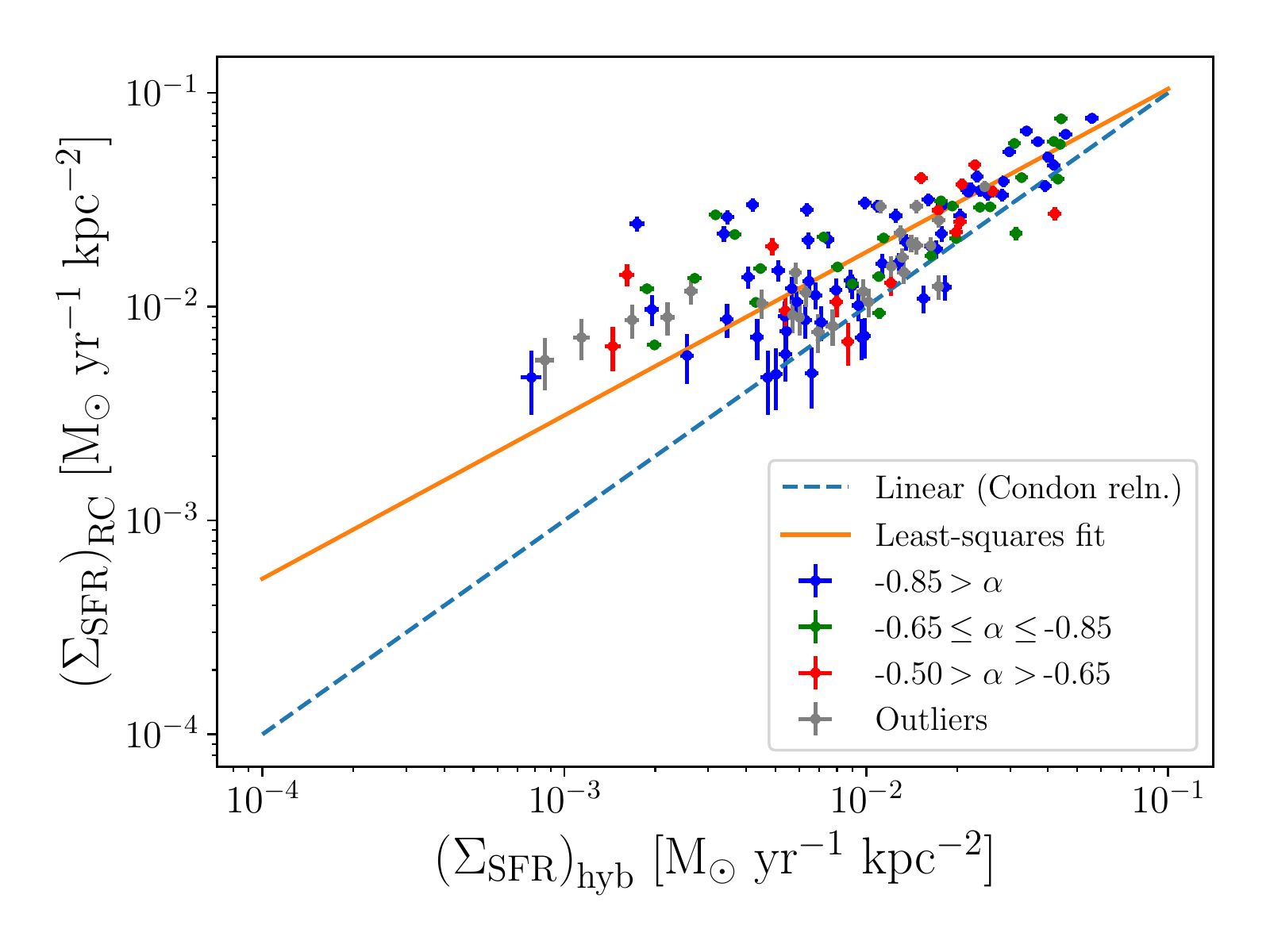}
    \end{subfigure}
    \begin{subfigure}[t]{0.03\linewidth}
        \textbf{(d)}    
    \end{subfigure}
    \begin{subfigure}[t]{0.47\linewidth}
            \includegraphics[width=\linewidth,valign=t]{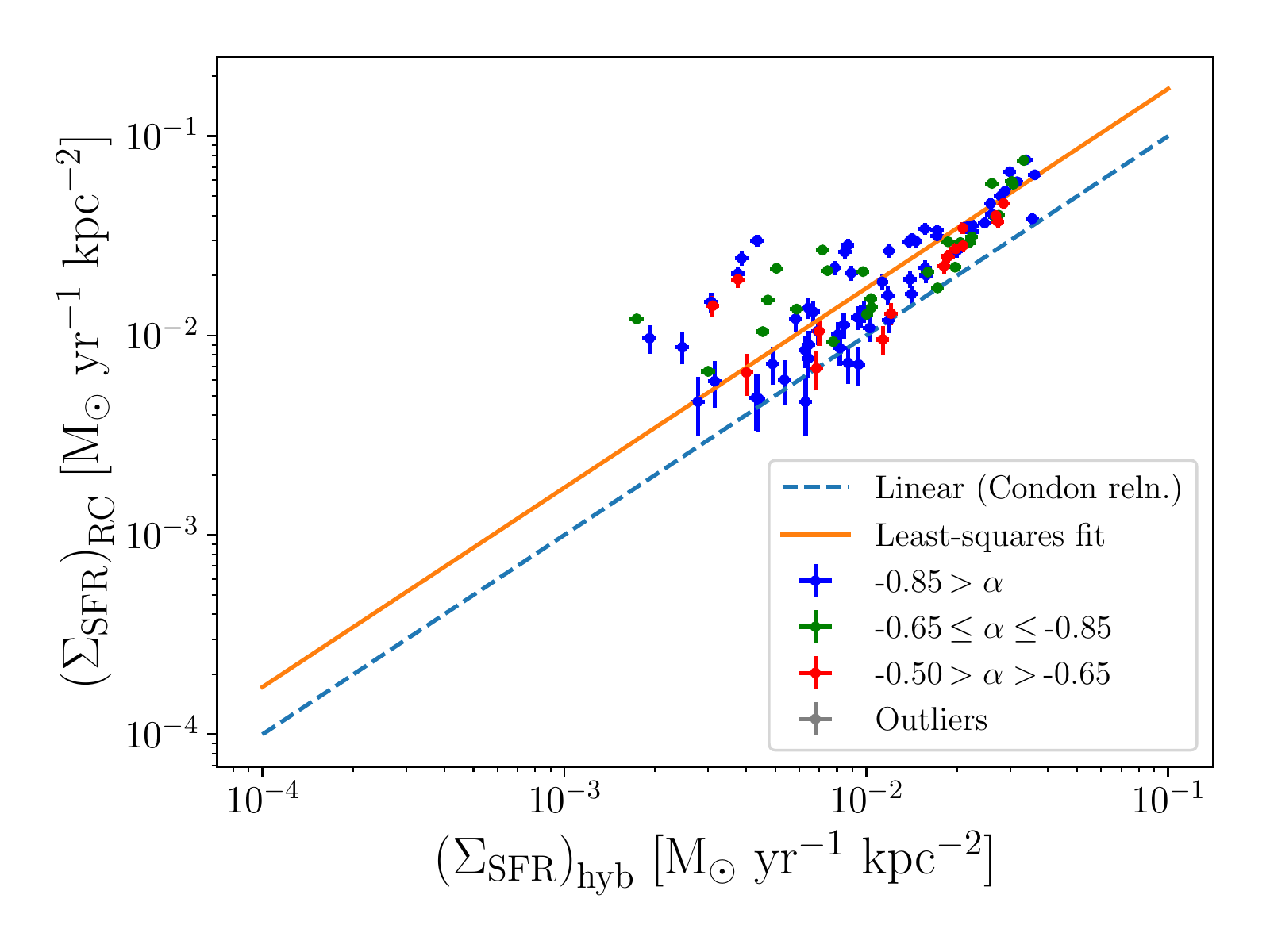}
    \end{subfigure}
    \caption{Figure~\ref{fig:pixel} continued. The top row shows plots at 4850~MHz (in combination with 8350~MHz, panels a and b), the bottom row shows plots at 8350~MHz (in combination with 4850~MHz, panels c and d).}
    \label{fig:pixel2}
\end{figure*}

\begin{table*}
	\centering
	\caption{Fitting results of the radio--SFR relation and its linearisation.}
	\label{tab:fitting}
	\begin{tabular}{cc cc cc cc cc cc} % four columns, alignment for each
		\hline
                $\nu_1$ & $\nu_2$ & $a_1$ & $b_1$ & $a_2$ & $b_2$ & $\sigma_1$  & $\sigma_2$ \\ 		
                \multicolumn{2}{c}{(MHz)} &        &        &  & & (dex) & (dex) \\
                (1)  & (2)        & (3)    & (4)               & (5)               & (6)              & (7)               & (8)    \\ \hline
                54   & 144        & $0.444\pm 0.015$  & $-0.934\pm 0.042$ & $0.617\pm 0.019$ & $-0.527\pm 0.052$ & $0.15$ & $0.20$ \\
                144  & 1365       & $0.518\pm 0.016$  & $-0.720\pm 0.042$ & $0.723\pm 0.017$ & $-0.172\pm 0.043$ & $0.19$ & $0.20$ \\ 	
                1365 & 4850       & $0.633\pm 0.041$  & $-0.348\pm 0.081$ & $0.787\pm 0.048$ & $-0.189\pm 0.097$ & $0.19$ & $0.23$ \\
                4850 &	8350      & $0.714\pm 0.053$  & $-0.292\pm 0.104$ & $0.764\pm 0.057$ & $-0.218\pm 0.110$ & $0.26$  & $0.28$ \\ 
                \hline
        \end{tabular}
        \flushleft
            {\small {\bf Notes.} Column (1) lower frequency for the spectral index; (2) higher frequency; (3+4) radio--SFR relation for the lower frequency; (5+6) same for the high frequency; (7+8) rms scatter.}
\end{table*}

\section{Results}
\label{s:results}

\subsection{Radio--SFR relation}

Our findings of the spatially resolved radio--SFR relation are in qualitative agreement with what is theoretically expected. Figure~\ref{fig:pixel} shows that the data points follow a sub-linear radio--SFR relation for all five frequencies. The slope of the relation is flatter for lower frequencies which is expected: we see older and therefore lower energy CREs at lower frequencies. We divide the points into three groups based on their radio spectral indices: young CREs ($-0.50 > \alpha > -0.65 $), medium age CREs ($-0.65 \geq \alpha \geq -0.85 $) and old CREs ($-0.85 > \alpha$). The values for the spectral indices are typical values found in the spiral arms, inter arm regions and outskirts of galaxies, respectively. Focusing just on the young CREs, the data points are fairly close to the Condon 1:1 relation. This behaviour is explained by the fact that the young CREs have energy spectra close to their injection spectra. They have not yet moved far from their sources and thus have not lost a lot of energy. That implies their spatial distribution should more closely resemble that of the star formation, they should therefore follow the Condon--relation more closely \citep{heesen_19a}.

%The best-fitting radio--SFR relations are presented in Table~\ref{tab:fitting}. We can also compare our results quantitatively to those in \textcite{Heesen2019} who used the same technique, but calculated the best fit slopes with linear regression instead of ODR. We argue that the $x$--errors in our data are not negligible and therefore the results obtained using ODR should be trusted more. A caveat here is that for ODR to work both errors have to either be symmetrical or both be normally distributed. It is hard to determine if this condition is satisfied for our data. That is why different measures to estimate the CRE transport lengths are also investigated using the method by \citet{murphy_08a} and \citet{vollmer_20a}. 

\begin{table*}
	\centering
	\caption{CRE transport length.}
	\label{tab:length}
	\begin{tabular}{cc cc c cc cc cc} % four columns, alignment for each
		\hline
                $\nu_1$ & $\nu_2$ &  $t_{\rm syn,1}$ & $t_{\rm syn,2}$ &FWHM  & $l_{\rm CRE,1}$  & $l_{\rm CRE,2}$ & $\sigma_{\rm l1}$ & $\sigma_{\rm l2}$ \\ 		
                \multicolumn{2}{c}{(MHz)}  & (Myr) & (Myr) & (kpc)  &  (kpc)     & (kpc) & (dex) & (dex) \\
                (1)  & (2)        & (3)& (4)& (5)    & (6)            & (7)            & (8)    & (9)   \\\hline
                54   & 144        & 89 & 55 & 1.8    & $5.23\pm 0.15$ & $3.70\pm 0.15$ & $0.15$ & $0.16$ \\
                144  & 1365       & 55 & 17.7 & 1.2    & $3.54\pm 0.15$ & $2.26\pm 0.15$ & $0.18$ & $0.16$ \\ 	
                1365 & 4850       & 17.7 & 9.4 & 1.2    & $2.52\pm 0.15$ & $1.47\pm 0.15$ & $0.18$ & $0.23$ \\
                4850 &	8350      & 9.4 & 7.2 & 1.2    & $1.74\pm 0.15$ & $1.34\pm 0.15$ & $0.20$ & $0.21$ \\ 
                \hline
        \end{tabular}
        \flushleft
            {\small {\bf Notes.} Column (1) lower frequency for the spectral index; (2) higher frequency; (3+4) CRE synchrotron lifetime at $\nu_1$ and $\nu_2$, respcetively; (5) angular resolution of map as full-width at half-maximum; (6+7) CRE transport length at $\nu_1$ and $\nu_2$, respectively; (8+9) rms scatter around linearised radio--SFR relation at $\nu_1$ and $\nu_2$, respectively.}
\end{table*}

\subsection{Cosmic-ray electron transport lengths}

The CRE transport lengths lie between $1.34$ and $5.23$~kpc and are thus smaller than the radius of the star-forming disc which is approximately 10~kpc \citep{mulcahy_14a}. Hence, the galaxy is sufficiently large to explore the transport length to these values. The cosmic-ray transport length decreases with frequency, which confirms that the length increases with lower CRE energy. For the frequencies where we have two measurements, we see differences in the transport length. These can be attributed to different radio spectral indices and data point numbers due to sensitivity limits. While the differences are larger than the statistical errors, they are still small (approximately 0.3~kpc).

We now calculate the electron lifetime taking into account synchrotron and inverse Compton losses. For that we use the average total (mostly random) equipartition magnetic field strength of $\langle B \rangle = 12.0\pm 2.6~\rm\upmu G$ \citep{heesen_22b}. The radiation energy density $U_{\rm rad} $ is the sum of two parts: the CMB energy density, which can be measured directly, and the total infrared energy density $U_{\rm TIR}$ with an additional contribution of stellar light. We used the total infrared luminosity of \citet{dale_09a} and used a radius of 16~kpc \citep[equivalent to the star-forming radius of][]{heesen_22a} to calculate $U_{\rm rad}=13\times 10^{-13}~\rm erg\,cm^{-3}$ \citep[see appendix~A in][]{heesen_18b}. With $U_{\rm B}=B^2/(8\uppi)$ we obtain $U_{\rm rad}/U_{\rm B}=0.23$ meaning synchrotron losses dominate over inverse Compton losses.

With these results we can write the energy and lifetime of the CREs as functions of parameters we know. Combining numerical factors and using convenient units yields:
\begin{equation}
    E({\rm GeV}) = \sqrt{  \left(\frac{\nu }{16.1~\rm MHz}\right) \left( \frac{B}{\rm \upmu G}\right)^{-1} } 
\end{equation}
and the electron lifetime:
\begin{equation}
t_{\rm syn} = 8.352\times 10^9 \left( \frac{E}{\rm GeV}\right)^{-1} \left( \frac{B}{\rm \upmu G}\right)^{-2} \left( 1 + \frac{U_{\text{rad}}}{U_{\text{B}}} \right)^{-1} ~\rm yr.
\end{equation}

\begin{figure}
	\includegraphics[width=\columnwidth]{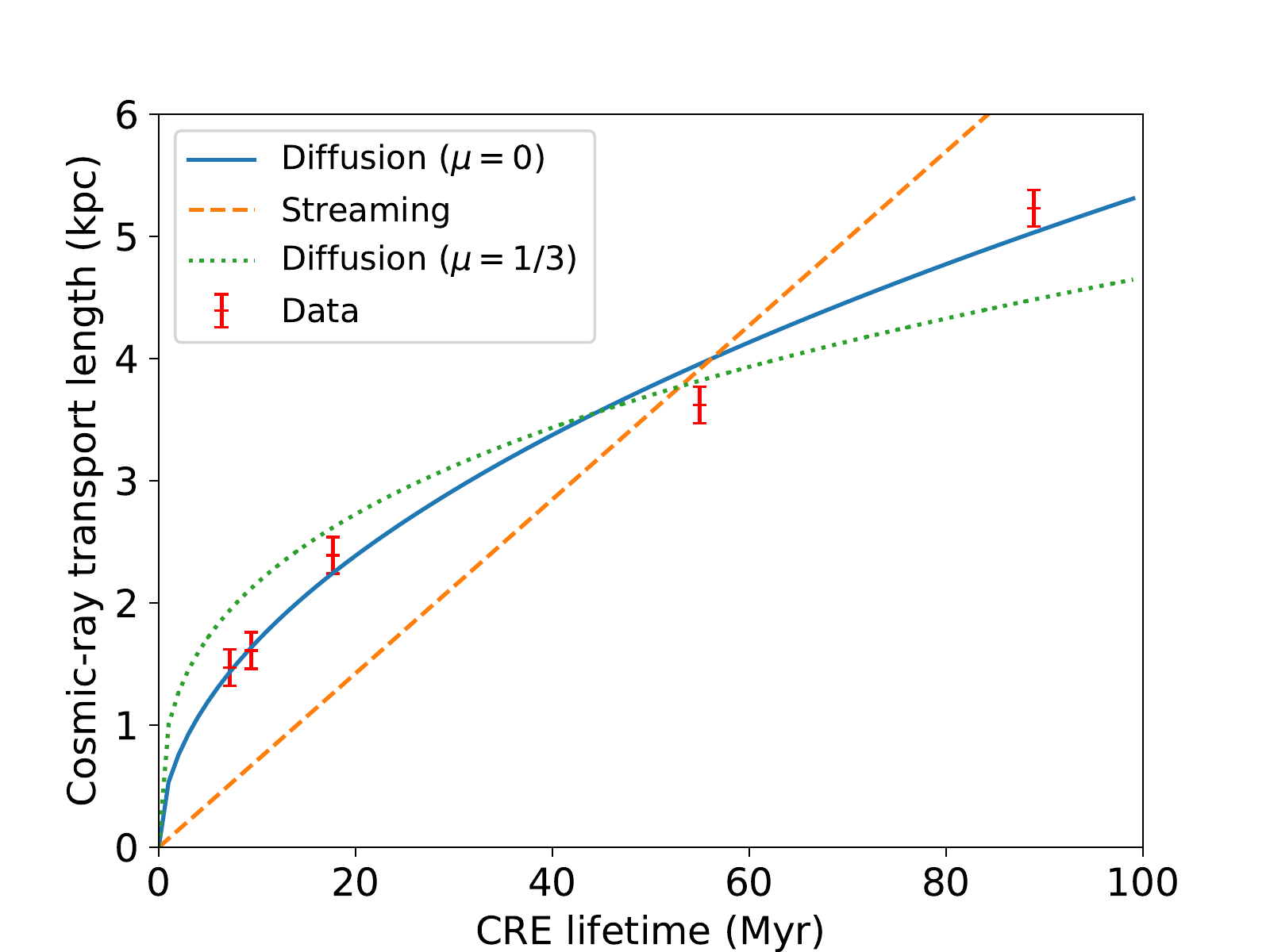}
    \caption{Cosmic-ray transport models. We show the transport length as function of CRE lifetime. The solid line shows the best-fitting diffusion model with energy independent diffusion; the dashed line shows the best-fitting streaming model; the dotted line shows the best-fitting diffusion model with an energy-dependent diffusion coefficient with $D\propto E^{1/3}$. Our data favours energy-independent diffusion.}
    \label{fig:diffusion}
\end{figure}

In Fig.~\ref{fig:diffusion}, we plot the cosmic-ray transport length as function of the electron lifetime. For cosmic-ray diffusion, we expect following dependence \citep{syrovatskii_59a}:
\begin{equation}
    l_{\rm CRE} = \sqrt{4Dt_{\rm syn}},
    \label{eq:diffusion}
\end{equation}
where $D$ is the isotropic diffusion coefficient. We find that Equation~\eqref{eq:diffusion} describes our data sufficiently well ($\chi_\nu^2=2.0$). The resulting diffusion coefficient is $(2.14\pm 0.13)\times 10^{28}~\rm cm^2\,s^{-1}$. This means that for the relevant energy range considered, $0.53$--$6.57$~GeV, the diffusion coefficient is energy-independent. This can be quantified further as the CRE energy and the lifetime are related as $t_{\rm syn}\propto E^{-1}$, so that by assuming an energy-dependence of $D\propto E^\mu$ we have following relation:
\begin{equation}
    l_{\rm CRE} \propto t_{\rm syn}^{0.5-\mu/2}.
    \label{eq:diff_energy_dependence}
\end{equation}
We thus find $\mu=-0.01\pm 0.20$ in agreement with an energy-independent diffusion coefficient.

We can rule out a linear dependence between transport length and lifetime $l_{\rm CRE}\propto t_{\rm syn}$, which would be the case for either cosmic-ray streaming or advection. This would result in a much steeper relation than what is observed in Fig.~\ref{fig:diffusion}. Similarly, energy-dependent diffusion coefficient would result in a flatter relation than what is observed. In Fig.~\ref{fig:diffusion}, we show the best-fitting relation with $D\propto E^{1/3}$, effectively meaning $l_{\rm CRE}\propto t_{\rm syn}^{1/3}$ (Equation~\ref{eq:diff_energy_dependence}), which is not a good fit to our data ($\chi_\nu^2=14.4$) when compared with the energy-independent case. Our diffusion coefficient is in good agreement with the value measured by \citet{heesen_19b}, who used the same method but only with two frequencies of 144 and 1365~MHz.

What we have neglected in our analysis thus far is the influence of CRE escape from the galaxy. Such an escape is required in order to explain the relatively flat radio continuum spectrum. Albeit being  clearly in agreement with non-thermal synchrotron emission it is not in agreement with an aged spectrum \citep{mulcahy_14a}. The recent work by \citet{doerner_22a} estimated this escape time-scale in the range 20--300~Myr where a positive gradient with a galactocentric radius is found. If we take an average value at a radius of 7~kpc, we would expect an escape time-scale of $t_{\rm esc}\approx 50~\rm Myr$ \citep[][their fig.~4]{doerner_22a}. We then calculated the effective CRE lifetime $1/t_{\rm CRE}=1/t_{\rm syn}+1/t_{\rm esc}$, assuming an exponential decay with time, and repeated the analysis. We found that neither diffusion nor streaming can adequately describe the data. Presumably the assumption of a constant escape time-scale is too simplistic for our data; at $\nu < 5$~GHz we expect long escape time-scales, whereas at $\nu>$5~GHz we have such short synchrotron loss-time scales so that escape does not play much of a role.

\section{Discussion}
\label{s:discussion}

The measurements presented in this paper deal with CREs, while lots of literature (both observations and theory) deal with cosmic-ray protons. 
Transport theories (e.g. quasi-linear theory and its non-linear extensions) of charged particles primarily rely on their gyro radius and the magnetic field they propagate through. Consequently, ultra-relativistic cosmic-ray protons and CREs of the same energy share the same transport properties in these theoretical transport models. 
However, one of the differences between electrons and protons in our GeV energy-range is that only electrons are ultra-relativistic, whereas protons are only mildly relativistic at best. It is however thus far unclear to what extent this would change the transport. Simulations modelling both protons and electrons with diffusion give good agreement with the local energy spectra near the Earth measured by {\it Voyager} of both protons and electrons \citep{werhahn_21a}. Similarly, $\gamma$-ray observations of nearby galaxies are usually described by diffusion of both electrons and protons \citep{peretti_19a}. Hence, while we have to add the disclaimer that electron and proton transport may be different, observations thus far allow us not to make a distinction and so in the following we may discuss cosmic-ray transport in general.

Particle experiments near Earth can directly detect the different components of the cosmic rays: electrons, positrons, protons and heavier nuclei. These are, however, limited to energies above 10 GeV because magnetic fields carried by the solar wind constrain the movements of low-energy cosmic rays. There is also a difference in the probed energy regime: the typical energy for CREs responsible for radio continuum emission is approximately 1--10 GeV, while observations in the solar system allow us to gain insights about the ion component which is inaccessible for radio observations. There is for instance a relative overabundance of light elements (He, Li, Be, B) by 5--7 orders of magnitude at $\approx 1$ GeV when compared to the interstellar medium. These can be accounted for if cosmic rays undergo spallation reactions with the interstellar medium \citep{zweibel_13a}, so it is possible to calculate the energy dependence of the diffusion coefficient:
\begin{equation}
    \frac{n_{\rm B}}{n_{\rm C}} \propto D^{-1} \propto E^{-1/3},
    \label{eq:energy_dependence}
\end{equation}
where $n_B/n_C$ is the Boron-to-Carbon ratio and the exponent is derived using Kolomogorov scaling analysis \citep{becker_tjus_20a}. The physical explanation for the energy dependence is that cosmic rays with higher energies and hence larger Larmor radii resonantly interact with fluctuations with larger wavelengths of the Kolmogorov spectrum leading to larger values of the mean free path and hence to higher diffusion coefficients.

In contrast, we showed that the most likely transport mechanism for CREs in our studied galaxy is energy-independent diffusion. In the following we discuss possible explanations.

The first theory that we consider suggests that the cosmic rays simply follow the magnetic field lines \citep{minnie_09a, reichherzer_22a}. Since the turbulence in the interstellar medium is super-Alfv{\'e}nic the field lines are bent and randomised too. Then we can, when viewed on scales larger than the coherence length $l_{\rm{c}}$ of the turbulence, describe the process of cosmic-ray transport by diffusion due to field line random walk (FLRW) in the turbulence of the ISM. While the particles follow the field lines they undergo spatial diffusion as the magnetic field lines diffuse in space over time. The resulting decorrelation of the particle trajectories \citep{jokipii_68a,hauff_10a} gives rise to an energy-independent mean free path and thus also an energy-independent diffusion coefficient. Such a picture is applicable if the cosmic-ray gyro radius is much smaller than the field coherence length,

\begin{equation}
\frac{r_{\rm g}}{l_{\rm{c}}}\approx 10^{-8} \left( \frac{E}{\rm GeV}\right) \left( \frac{B}{10~\rm \upmu G}\right)^{-1} \left( \frac{l_{\rm{c}}}{10~{\rm pc}}\right)^{-1}.
\end{equation}
 
Calculations show that this mechanism is accurate for cosmic rays with energies of up to 100 GeV. For cosmic rays at even higher energies another explanation is needed for their measured isotropy; however, \citet{reichherzer_22a} showed that the anisotropic diffusion in isotropic Kolmogorov turbulence for such particles may still be described by FLRW.

The second theory is cosmic ray self-confinement. Cosmic rays create their own turbulence via waves that are generated by the cosmic ray streaming instability \citep{kulsrud_69a,zweibel_13a}. The cosmic rays are then confined by interactions with small-scale fluctuations in the magnetic field for tens of Myr. Due to this confinement and frequent scattering with the magnetic field, their distribution becomes isotropic and their large-scale motions can accurately be described by energy-independent transport. The cosmic rays stream at the Alfv{\'e}n speed \citep{skilling_71a}, and so the cosmic ray transport is energy-independent. However, in the presence of wave damping, cosmic rays may stream faster than at Alfv{\'e}n speed depending on energy, so that the transport is not energy-independent anymore. Also, the growth rate of the turbulence is energy dependent as well \citep{farmer_04a}. That said \citet{krumholz_20a} still suggest that the cosmic rays stream along magnetic files lines with a speed close to the Alfv{\'e}n speed as long as they are decoupled from the neutral gas. A similar result for a fully ionised gas was found by \citet{yan_08a}.

The third theory is scattering of cosmic rays in the turbulent magnetic field, where the source of turbulence is extrinsic. Turbulence is injected at scales of $\approx$50~pc and cascades down to the resonant scale of the gyro radius. Simple analytic models of such a transport predicts an energy-dependent diffusion coefficient, where the slope depends on the assumed turbulence model. For instance, the standard assumption is $D\propto E^{1/3}$ as expected for Kolmogorov-type turbulence \citep{strong_07a}. But as \citet{kempski_22a} calculate, the scattering at magnetohydrodynamic fast-mode waves of low energy cosmic rays in the warm ISM results in energy-independent diffusion. Notably their calculated diffusion coefficients of around $2\times 10^{28} {\rm cm^2\,s^{-1}}$ \citep[see fig.~3 in][]{kempski_22a} is in good agreement with our result.

Unfortunately, the regime of turbulence most relevant to cosmic ray transport of GeV electrons is difficult to access in numerical simulations, so that there is no confirmation of either theory from that direction. The reason is that due to the Kolmogorov scaling of the external turbulence, there is little energy in the fluctuations available for resonant scattering of the sub-TeV CREs. This is in contrast to the regimes investigated in test-particle simulations \citep[e.g.][]{Giacalone_99, Subedi_17, reichherzer_22a, Kuhlen_22}, where only parameter spaces above $r_{\rm g}/l_{\rm{c}} \approx 10^{-4}$, corresponding to energies $\gtrsim 10\,$TeV, were investigated and for which energy-dependent diffusion is found.

Our approach of measuring CRE transport by comparing the radio continuum map with the SFR surface density is complementary to the radio spectral index analysis by \citet{mulcahy_16a} and \citet{doerner_22a}. While our method is sensitive to smoothing of the \sfrd-map by CRE diffusion, the radio spectral index distribution is more governed by the escape of CRE from the galaxy. Our best-fitting diffusion coefficient of $D\approx 2\times 10^{28}$~\udif\ is in good agreement with the results of \citet{doerner_22a} but lower than the value by \citet{mulcahy_16a} who found $D\approx 6\times 10^{28}$~\udif. The difference can be mostly attributed to the fact that \citet{mulcahy_16a} modelled CRE escape only via diffusion, whereas \citet{doerner_22a} used a combination of diffusion and advection. Consequently, the diffusion coefficient has to be higher in order to allow for a fast enough CRE escape \citep{mulcahy_16a}. Our result shows that a lower diffusion coefficient is in agreement both with the smoothing experiment and the radio spectral index analysis, when the effect of a galactic wind is included in the latter.

\section{Conclusions}
\label{s:conclusions}

The transport of cosmic rays in the nearby galaxy M~51 is studied using ultra-low frequency data at 54~MHz from the LoLSS survey \citep{de_gasperin_21a}. We complement these data with radio continuum maps at 144~MHz from LoTSS-DR2 \citep{shimwell_19a}, 1365 MHz from WSRT--SINGS \citep{braun_07a}, and 4850 and 8350~MHz from merged VLA and Effelsberg observations \citep{fletcher_11a}. In order to measure spatially resolved star formation, we use a hybrid \sfrd-map from a combination of {\it Spitzer} 24-$\upmu$m and {\it GALEX} 156-nm emission from the THINGS survey \citep{leroy_08a}. We correct for thermal radio continuum using a combination of H\,$\alpha$ and {\it Spitzer} infrared observations. These are our main conclusions:
\begin{itemize}
    \item The radio continuum maps appear significantly smoothed out in comparison with the \sfrd-maps. This can be quantified using the spatially resolved radio--SFR relation at 1.2~kpc resolution. The slopes of the radio--SFR relation are sub-linear meaning an excess of radio continuum emission at low values of \sfrd and underluminous radio continuum at high values of \sfrd. As the global (integrated) radio--SFR relation in galaxies is approximately linear \citep[e.g.][]{smith_21a}, this can be understood as the result of cosmic-ray transport.
    \item The slopes of the radio--SFR relation are frequency dependent, where the lowest frequencies have the lowest slopes. This is expected, as the CRE lifetime increases for the lowest frequencies and so the CRE transport length increases as well which amplifies the effect of the smoothing; this then lessens the relation between radio continuum emission and star formation.
    \item We smooth the \sfrd-map with a Gaussian kernel in order to mimic cosmic-ray diffusion. Thus we are able to linearise the spatially resolved radio--SFR relation. The CRE transport length is then half of the FWHM of the smoothing Gaussian function. The CRE transport length decreases with increasing frequency as expected since the CRE lifetime decreases as well.
    \item When we plot the CRE transport length as function of the CRE lifetime (Fig.~\ref{fig:diffusion}), we find a square-root dependence exactly as expected for diffusion. When we fit our data as $l_{\rm CRE}=\sqrt{4Dt_{\rm syn}}$, we can derive the isotropic diffusion coefficient of $D=(2.14\pm 0.13)\times 10^{28}~\rm cm^2\,s^{-1}$. 
    \item The diffusion coefficient is with $D\propto E^{0.001\pm 0.185}$ energy-independent across the relevant energy range considered, $0.53$--$6.57$~GeV.  This result requires some theoretical explanation, such as FLRW, cosmic ray self-confinement or cosmic ray scattering at extrinsic turbulence.
\end{itemize}
One outcome of our work is that the radio spectral index information is useful to identify data points which are dominated by thermal emission or suffer from thermal absorption, or, possibly, a combination of both. In the future, we may be able to correct more accurately for thermal emission and absorption by modelling radio continuum spectra exploiting the ultra-low frequency emission from LoLSS \citep{de_gasperin_21a}. Also, spatially resolved optical spectral with integral field unit spectroscopy allows one now to obtain extinction-corrected H\,$\alpha$ maps by using the theoretical H\,$\alpha$/H\,$\beta$ ratio. This is now performed for THINGS galaxies as part of the Metal--THINGS survey \citep{lara_lopez_21a}.

\begin{acknowledgement}
We thank the anonymous referee for an insightful and stimulating report. We furthermore thank Andrew Fletcher for useful inputs to the manuscript. We also wish to thank Philipp Grete for a fruitful discussion. This paper is based (in part) on data obtained with the International LOFAR Telescope (ILT). LOFAR \citep{vanHaarlem_13a} is the Low Frequency Array designed and constructed by ASTRON. It has observing, data processing, and data storage facilities in several countries, that are owned by various parties (each with their own funding sources), and that are collectively operated by the ILT foundation under a joint scientific policy. The ILT resources have benefitted from the following recent major funding sources: CNRS-INSU, Observatoire de Paris and Universit\'e d'Orl\'eans, France; BMBF, MIWF-NRW, MPG, Germany; Science Foundation Ireland (SFI), Department of Business, Enterprise and Innovation (DBEI), Ireland; NWO, The Netherlands; The Science and Technology Facilities Council, UK; Ministry of Science and Higher Education, Poland.

M.B. acknowledges funding by the Deutsche Forschungsgemeinschaft (DFG, German Research Foundation) under Germany’s Excellence Strategy – EXC 2121 ‘Quantum Universe’ – 390833306. RJD and PR also acknowledge DFG funding via the Collaborative Research Center SFB1491 "Cosmic Interacting Matters" - 445052434. This work made use of the {\sc SciPy} project \href{https://scipy.org}{https://scipy.org}. We use the colour scheme of \citet{green_11a} to display the radio spectral index maps.

\end{acknowledgement}


\bibliographystyle{aa}
\bibliography{review} % if your bibtex file is called example.bib

\end{document}